\documentclass[12pt]{article}
\usepackage{graphicx}
\usepackage{amsmath}
\usepackage{amsfonts}
\usepackage{amssymb}    
\usepackage{graphicx}   
\usepackage{cite}    
\newcommand{\be}{\begin{equation}}
\newcommand{\ee}{\end{equation}}
\newcommand{\bea}{\begin{eqnarray}}
\newcommand{\eea}{\end{eqnarray}}
\newcommand{\ba}{\begin{array}}
\newcommand{\ea}{\end{array}}

\newcommand{\no}{\nonumber}

\newcommand{\cO}{{\cal O}}
\newcommand{\cA}{{\cal A}}
\newcommand{\cM}{{\cal M}}
\newcommand{\GG}{\Omega}

\begin{document}

\begin{flushright}
July 2007 \\
\end{flushright}
\vskip   1 true cm 
\begin{center}
{\Large \textbf{Soft-photon corrections in multi-body meson decays}}    \\ [20 pt]
\textsc{Gino Isidori}   \\ [20 pt]
\textsl{INFN, Laboratori Nazionali di Frascati, I-00044 Frascati,Italy} 

\vskip   1 true cm 

\textbf{Abstract}
\end{center}
\noindent  
The effects due to soft-photon emission (and the related virtual corrections)
in multi-body decays of $B$, $D$, and $K$ mesons are analysed. 
We present analytic expressions for the  universal $\cO(\alpha)$ correction 
factors which can be applied to all multi-body decay modes where a tight soft-photon 
energy cut in the decaying-particle rest-frame is applied. 
All-order resummations valid in the limit of small and large 
velocities of the final-state particles are also discussed. 
The phenomenological implications of these correction 
factors in the distortion of Dalitz-plot distributions of 
$K\to 3\pi$  decays are briefly analysed. 

\vskip   1 true cm 
          
%%\paragraph{I.}

\section{Introduction}
In the last few years the large amount of data collected 
at flavour factories has allowed to reach statistical accuracies 
around or below the percent level in several 
decays modes of $B$, $D$, and $K$ mesons.  At this level of accuracy 
electromagnetic effects cannot be neglected. The theoretical 
evaluation of these effects is a key ingredient  
to extract from data a precise information about 
weak interactions or strong dynamics,
such as the determination of CKM matrix 
elements or the extraction of 
of $\pi\pi$ scattering lengths.

The theoretical treatment of the infrared singularities 
generated within QED is a well known subject and one 
of the pillars of quantum field theory. A clear 
and very general discussion can be found, for instance, 
in Ref.~\cite{Yen,Wei}. 
These general properties of QED have been exploited 
in great detail in the case of genuine electroweak 
processes, or processes which can be fully described 
within perturbation theory within the Standard Model (SM).
More recently, a similar program has been extended 
to a few decay modes of $K$ and $B$ mesons 
(see e.g.~Ref.~\cite{Cir,Kl3,Bijnens,Gatti,BPP}),
which can be described within appropriate 
effective field theories (EFT). The purpose of the present article 
is to complement and generalise these EFT studies, analysing the general 
structure of electromagnetic corrections in multi-particle 
final states. In particular, we are interested in the 
distortions of the non-radiative decay distributions (Dalitz plot parameters, 
form factor slopes, etc\ldots) induced 
by electromagnetic effects. To a large extent,
these effects have a {\em universal} (long-distance) character:
their structure can be evaluated independently of 
the short-distance dynamics which originate the meson decay.

%%\paragraph{II.}

\section{The photon-inclusive decay distribution at $\cO(\alpha)$}
From the experimental point of view, 
the most convenient infrared-safe observable related to the process
$P_0 \to P_1 \ldots P_N$ is the differential photon-inclusive
distribution 
\be
{\rm d} \Gamma^{\rm incl}(s_{ij}; E^{\rm max})  = \left. {\rm d} 
\Gamma (P_0 \to P_1 \ldots P_N + n\gamma)\, \right|_{\sum E_{\gamma} < E^{\rm max} }~,
\label{eq:dgincl}
\ee
namely the differential width for the process $P_0 \to P_1 \ldots P_N$ accompanied 
by any number of (undetected) photons, with total missing energy less or equal to 
$E^{\rm max}$  in the $P_0$ rest frame. In addition to $E^{\rm max}$,
the differential photon-inclusive distribution depends on kinematical variables 
describing the visible particles. A convenient choice for the latter 
is\footnote{~Although redundant, this choice of variables
allows us to keep the discussion on a general ground.}
\be
s_{ij} = \left\{ \ba{ll} (p_i +p_j)^2 \qquad & i\not= 0,~j\not= 0,~  \\  (p_0 -p_j)^2 &  i=0,~j\not= 0. 
\ea \right. 
\ee

The photon-inclusive distribution in (\ref{eq:dgincl}) can be decomposed
as the product of two theoretical quantities: the so-called 
non-radiative width, ${\rm d}\Gamma^0(s_{ij})$, which survives 
in the $\alpha\to 0$ limit, and the corresponding energy-dependent 
electromagnetic correction factor $\GG(s_{ij}; E^{\rm max})$:
\be
{\rm d} \Gamma^{\rm incl}(s_{ij}; E^{\rm max})  = 
{\rm d} \Gamma^{\rm 0}(s_{ij}) \times  \GG(s_{ij}; E^{\rm max})~.
\label{eq:prod}
\ee
At any order in the perturbative expansion in $\alpha$
the energy dependence of  $\GG(s_{ij}; E)$ is unambiguous 
and universal  up to terms which vanish in the 
limit $E \to 0$ \cite{Wei}. The $E$-independent part of 
$\GG$ contains both universal terms, such as the Coulomb 
corrections, and non-universal terms depending on the 
short-distance dynamics which originate the decay.
In order to discuss the separation between universal and 
non-universal terms, we start presenting the calculation 
of $\GG(s_{ij}; E)$ at $\cO(\alpha)$ in the limit of 
a real point-like effective weak vertex. 

The general decomposition of $\GG(s_{ij}; E)$ at  $\cO(\alpha)$ is
\bea
\GG(s_{ij};  E)  &=& 1+ \sum_{i,j=0}^N Q_i Q_j J_{ij}(s_{ij};  E)~, \\
 J_{ij}(s_{ij};  E) 
&=& \frac{\alpha}{\pi} \left[ 2 b_{ij} \ln\left( \frac{m_0}{2 E} \right) 
 + F_{ij} + H_{ij}^{\rm IR} +  H_{ij}^{\rm C} + H_{ij}^{\rm UV} 
 + \cO\left( E \right)\,
\right]~,
\label{eq:G_1}
\eea
where $Q_{i\not=0}$ are the charges of the final-state particles 
in units of $e$ and $Q_{0}=-\sum_{i=1}^N Q_i$. The terms $b_{ij}$ and $F_{ij}$ 
are unambiguously determined by the real-emission amplitude, 
while the $H_{ij}$ functions are associated to virtual corrections. 

The $S$ matrix element corresponding to the emission of 
a real photon can be decomposed as 
\be
 _{\rm out} \langle  P_1(p_1) \ldots P_N(p_n) + \gamma(\epsilon,k)
| P_0 (p_0) \rangle_{\rm in}  = - i e \cM_0 \times \hat{K} \times 
 (2\pi)^4 \delta^4(p_0 - \sum_{i=1}^N p_i -k)~, 
\label{eq:real_1} 
\ee
where $\cM_0$ is the invariant amplitude of the non-radiative process:
\be
 _{\rm out} \langle  P_1(p_1) \ldots P_N(p_n) 
| P_0 (p_0) \rangle_{\rm in}  = - i \cM_0~ (2\pi)^4 \delta^4(p_0 - \sum_{i=1}^N p_i)
\ee
and
\be
\hat{K} = 
\sum_{i=0}^N Q_i \frac {\epsilon \cdot p_i }{k \cdot p_i} +\cO(k)~.
\ee
The integration of the real-emission amplitude in the soft-photon 
approximation with a photon-energy cut $E$ 
(namely neglecting $\cO(E)$ terms) and regularizing the 
infrared-singularities with a photon mass $m_\gamma$, leads to 
\bea
{\rm d} \Gamma^{\rm real}(s_{ij}; E)  &=& 
{\rm d} \Gamma^{\rm 0}(s_{ij}) \times 
\int_{ E_\gamma <E} 
\frac{d^3 \vec{k} }{(2\pi)^3 ~2 E_\gamma}~ 
\sum_{\rm spins} | \hat K |^2 = \no\\
&=& {\rm d} \Gamma^{\rm 0}(s_{ij}) \times  \frac{\alpha}{\pi} \sum_{i,j=0}^N Q_i Q_j
\left[ 2 b_{ij} \ln\left( \frac{m_\gamma}{2 E} \right) 
 + F_{ij}  +\cO(E)\, \right]~,
\label{eq:real_2}
\eea
where \cite{Yen,Wei}
\be
b_{ii} = \frac{1}{2}~,  \qquad 
b_{i\not=j} =  \frac{1 }{ 4 \beta_{ij} } 
\ln \left( \frac{1 + \beta_{ij} }{1 - \beta_{ij}} \right)~, \qquad 
\beta_{ij} = \left[ 1 - \frac{ 4 m^2_i m^2_j }{ ( s_{ij} - m^2_{i} -m_{j}^2 )^2} \right]^{1/2}~.
\quad
\ee
The finite term $F_{ij}$ depends on the specific cut applied on the 
(soft) photon energy. Imposing the condition 
$p_0 \cdot k < m_0 E$, corresponding to a cut in the $P_0$ rest frame,
leads to 
\be
F_{i\not=j} = \Delta_{ij} \int_{-1}^{1} {\rm d}z~ \frac{ e(z) }{ p(z) 
[ e^2(z) -p^2(z) ]  }  \ln \left( \frac{e(z)+p(z) }{e(z) -p(z)} \right)~,
\ee
where
\bea
&& e(z) = \left(\frac{m_i m_j}{s_{ij}}\right)^{1/2}~ [\gamma_{0i} (1-z) + \gamma_{0j} (1+z)]~, 
\qquad\qquad\quad \gamma_{ij} = \frac{1}{ (1-\beta^2_{ij})^{1/2} }~, 
\\
&& p(z) = \left\{ \frac{m_i m_j}{s_{ij}} \left[(\gamma_{0i}^2-1) (1-z)^2 + (\gamma_{0j}^2-1)(1+z)^2 
\right] \right. \no \\
&& \qquad\quad \left.
+2\left( \gamma_{0i} \gamma_{0j} \frac{m_i m_j}{s_{ij}} - \Delta_{ij} \right)
 (1+z)(1-z) \right\}^{1/2}, \qquad 
\Delta_{ij} = \frac{s_{ij} - m^2_{i} -m_{j}^2 }{2 s_{ij} }~, \qquad\
\eea
with the special case $i=j$ given by 
\be
F_{ii}  = \frac{1}{2\beta_{0i}} \ln\left( \frac{1+\beta_{0i}}{1-\beta_{0i}} \right)~, \qquad 
F_{00} = 1~.
\ee

As far as virtual corrections are concerned, the universal infrared 
singular term cancels out the $\ln(m_\gamma)$ dependence in 
Eq.~(\ref{eq:real_2}), and the remaining finite terms are encoded 
into the three $H_{ij}$ functions in Eq.~(\ref{eq:G_1}). Regularizing 
UV divergences by means of dimensional regularization and renormalizing 
the real point-like weak vertex in the $\overline{\rm MS}$ scheme leads 
to
\bea
&& H_{ij}^{\rm C} = -  \frac{ \pi^2}{2 \beta_{ij}} 
 (1-\delta_{ij})~\Theta\left( \sqrt{s_{ij}} -m_i -m_j \right)~,
 \\
&& H_{ij}^{\rm UV} = \frac{1}{4} \ln\left(\frac{\mu^2}{m_0^2}\right) (-1+3\delta_{ij})~, \\
&& H_{ij}^{\rm IR} = (1-\delta_{ij}) \left\{ -\frac{1}{2} 
+ \frac{1}{4}\ln\left( \frac{s_{ij}}{m_0^2}  \right)
-\frac{ m_i^2 - m_j^2 }{4 s_{ij}} \ln\left( \frac{m_i}{m_j} \right)
+ \frac{1}{4}\ln\left( \frac{m_i m_j}{s_{ij}}  \right) \right.  \no \\
&& \quad\  
- \frac{1}{4} \beta_{ij} \Delta_{ij} \ln\left( \frac{1 + \beta_{ij} }{1 - \beta_{ij} } \right) 
+ \frac{1}{4 \beta_{ij}} \ln\left( \frac{ s_{ij} \beta_{ij}  |\Delta_{ij}| }{ m_0^2} \right)
 \ln\left( \frac{1 + \beta_{ij} }{1 - \beta_{ij} } \right) \no \\
&& \quad\ \left.
+ \frac{1}{8\beta_{ij}} \left[ f\left( \frac{\Delta_{i} -\Delta_{ij} \beta_{ij}}{\Delta_{i} +\Delta_{ij} \beta_{ij}}\right) + 2 \ln\left(  \frac{ s_{ij} \beta_{ij}  |\Delta_{ij}| }{ m_i^2} \right)
\ln\left( \frac{\Delta_{i} -\Delta_{ij} \beta_{ij}}{\Delta_{i} +\Delta_{ij} \beta_{ij}}\right)
+(i\leftrightarrow j) 
\right] \right\}, \quad\ \no \\
\label{eq:17}
\eea
where 
$$
\Delta_{i} = \frac{s_{ij} + m^2_{i} -m_{j}^2 }{2 s_{ij} }~, \quad
f(x) = - 4 \int_{0}^{x} {\rm d}t \frac{  \ln(1-t)}{t} + \ln^2(x)~.
$$

The first term, $H_{ij}^{\rm C}$, which is 
singular in the limit of vanishing velocity among any pair 
of charged particles, is a genuine long-distance effect: 
it corresponds to the Coulomb interaction among 
the two charged particles. This term can indeed be 
evaluated also in non-relativistic quantum mechanics 
by means of semi-classical methods (see e.g.~Ref.~\cite{Semicl}).

The second term,  $H_{ij}^{\rm UV}$, which depends explicitly 
on the ultraviolet renormalization scale $\mu$, is 
manifestly not universal: its scale dependence
cancels out in Eq.~(\ref{eq:prod}), or in the physical 
observable, by the corresponding scale dependence of the weak 
amplitude. The finite $\cO(\alpha)$ term resulting after this 
cancellation cannot be computed without knowing the  
short-distance behaviour of the amplitude. Note that, 
in the approximation of a point-like weak vertex, 
this missing piece affects only the overall
normalization of the photon-inclusive distribution
and not its kinematical structure. 
 
By construction, $H_{ij}^{\rm IR}$ is what remains after 
isolating the manifestly universal and manifestly non-universal 
terms $H_{ij}^{\rm C}$ and $H_{ij}^{\rm UV}$. 
More explicitly, $H_{ij}^{\rm IR}$ is the finite part 
of the universal three-point function function after 
subtracting ultraviolet and infrared divergences 
and the Coulomb term:\footnote{~The result in Eq.~(\ref{eq:17}) 
is is valid only for $s_{ij}$ variables 
in the physical range, namely $s_{ij}$ real and positive
(such that all terms in Eq.~(\ref{eq:17}) are real).} 
\bea 
H_{ij}^{\rm IR} &=& 4 \pi^2 (1-\delta_{ij})~\Re\left\{ 
\int_{{\overline {\rm MS}}}
 ~\frac{ d^d k }{ i(2\pi)^d} ~\frac{(2p_i +k)_\mu (2p_j-k)^\mu }{
\left[(p_i+k)^2 -m_i^2\right]\left[(p_j-k)^2 -m_j^2\right]\left[k^2 -m_\gamma^2\right] }  \right\} \no \\
&& - H_{ij}^{\rm C}  +(1-\delta_{ij}) \left[ \frac{1}{4} \ln\left( \frac{\mu^2}{m_0^2}\right)  
 + b_{ij}  \ln\left( \frac{m_\gamma^2}{m_0^2}\right) \right] ~.
\label{eq:18}
\eea

\section{Resummations and universal correction factor}
The $E\to 0$ singular terms in Eq.~(\ref{eq:G_1})
and the $\beta_{ij} \to 0$ singular terms in $H_{ij}^{\rm C}$,
which represent the potentially largest correction factors, 
can be summed to all orders in $\alpha$.

As shown in Ref.~\cite{Wei},  
the resummation of the $\alpha^n \ln^n(E)$ terms
allows us to remove the $E \to 0$ singularity, 
giving rise to the following exponential term
\be
\GG_{\rm B} (s_{ij};  E) = \left( \frac{2 E}{m_0} \right)^{\frac{2\alpha}{\pi} B(s_{ij})}~, 
\qquad B(s_{ij}) =  - \sum_{i,j=0}^N Q_i Q_j b_{ij} ~ > 0 ~.
\label{eq:Omega_B}
\ee 

The resummation of the $(\pi\alpha/\beta_{ij})^n$ Coulomb terms is encoded 
by the semi-classical result \cite{Semicl}
\be
\GG_{\rm C}(s_{ij}) = \prod_{ \{0< i < j\} }  ~ \frac{ 2 \pi \alpha Q_i Q_j}{ \beta_{ij} }
\frac{1}{ e^{  \frac{ 2 \pi \alpha Q_i Q_j}{ \beta_{ij} } } -1 }~=~ 
1 + \frac{\alpha}{\pi} \sum_{ij=0}^N Q_i Q_j H_{ij}^{C} + \cO(\alpha^2)~.
\label{eq:Omega_C}
\ee
The $\beta_{ij} \to 0 $ singularity does not disappear and it is strengthened
in the case of opposite-sign charges (attractive interaction), 
but it remains an integrable singularity over 
the final-state phase space.

The two resummed expressions in Eqs.~(\ref{eq:Omega_B}) and (\ref{eq:Omega_C})
are relevant in two different kinematical regimes: 
$\GG_{\rm C}(s_{ij})$ is relevant in the $\beta_{ij} \to 0$ limit,
while $\GG_{\rm B}(s_{ij};  E)$ acquires a non-trivial kinematical 
dependence only in the $\beta_{ij} \to 1$ limit.
We can therefore factorize the two effects up to sub-leading 
$\cO(\alpha^2)$ corrections. This allows us to consider the following 
generalization for the universal part of the electromagnetic 
correction factor:
\be
 \GG_{\rm eff}(s_{ij}; E) = \GG_{\rm B}(s_{ij};  E) \times \GG_{\rm C}(s_{ij})
\times \left[ 1 + \frac{\alpha}{\pi} \sum_{i,j=0}^N Q_i Q_j ( F_{ij} + H^{\rm IR}_{ij} )  \right]~.
\label{eq:main}
\ee

This expression provides a good description of the leading kinematical corrections 
induced by soft photons in multi-body meson decay.
The approximations/validity-limits of $ \GG_{\rm eff}(s_{ij}; E)$ can be listed 
as follows:
\begin{itemize}
\item 
The leading kinematical singularities, namely 
the $\alpha^n/\beta_{ij}^n$ terms for $\beta_{ij} \to 0$ 
and the $\alpha^n \ln^n(E/m_0)^n \ln^n(1-\beta_{ij})$ terms for $\beta_{ij} \to 1$,
are summed to all orders. 
\item 
The regular contribution of the 
real photon emission ($F_{ij}$) is correct up to constant terms of 
$\cO(\alpha^2)$ and energy-dependent terms of $\cO(\alpha E/\Lambda)$,       
where $\Lambda$ is a typical hadronic scale. More precisely, the 
corrections linear in $E$ are controlled by the derivatives 
of the non-radiative amplitude with respect the kinematical 
variables: $\cO(\alpha E \times \partial \cA/\partial s_i )$~\cite{4mrad}.  
In several cases the tightness on the photon-energy cut necessary 
to keep these corrections under control can thus be 
quantitatively controlled by the smoothness 
of the non-radiative amplitude. In practice, the photon-energy 
cut is rarely a problem in $\pi$ and $K$ decays\footnote{~The only exceptions 
are modes where the bremsstrahlung is strongly suppressed compared to 
the the direct emission by
symmetry arguments, such as the helicity-suppressed $K\to e\nu (\gamma)$ 
or the CP-violating  $K_L \to \pi^+\pi^-(\gamma)$.}, 
while it is a non-trivial constraint for heavier mesons.
\item  
The virtual corrections encoded in $H_{ij}^{\rm IR}$ are only the 
universal contribution of low-energy photons within an effective 
theory valid below the scale $\Lambda$ ($\Lambda < M_{\rho}$),
with real effective couplings in the $\alpha\to 0$ 
limit.
High-energy modes provides in general additional infrared-safe  $\cO(\alpha)$
contributions which should be evaluated mode by mode (non-universal terms),
and which are different in case of final-state leptons 
or mesons.\footnote{~Having assumed real effective couplings 
in the $\alpha\to 0$ limit, we have also ignored the $\cO(\alpha)$ 
electromagnetic corrections to the strong phases of the amplitude. 
For smooth strong phases these can be easily be incorporated starting from the 
imaginary part of the three-point function in Eq.~(\ref{eq:18}), 
as discussed for instance in Ref.~\cite{Cir} for the $K\to\pi\pi$ case.}
By an appropriate matching procedure, these additional terms can be 
reabsorbed into the normalization and the kinematical dependence 
of the non-radiative amplitude. In light meson ($\pi$ and $K$) decays 
these extra terms are necessarily smooth 
functions of the kinematical variables  $\cO(\alpha s_i /\Lambda^2)$ and thus 
can be safely neglected. These ultraviolet effects are potentially larger in 
heavy meson decays, but also in this case they are subleading with respect 
to the leading logarithmic singularities included 
in $\GG_{\rm eff}(s_{ij}; E)$. 
\item
The only cases where virtual effects not included in Eq.~(\ref{eq:main})
are potentially relevant are the singular points corresponding to the 
formation of Coulomb bound states. A notable example is the 
pionium formation~\cite{pionium},
which has recently been observed in $K \to 3 \pi$ decays~\cite{NA48a,NA48b}.
Such states are treated here as different final states, which 
should be eliminated by appropriate kinematical cuts (as 
done for instance in Ref.~\cite{NA48a,NA48b}). Given the extremely narrow 
widths of Coulomb bound states, and the low probability formation,  
these effects are relevant only in very tiny regions of the space space 
and can be safely neglected in heavy-meson decays.
\end{itemize}

\section{A specific application: $K^+ \to \pi^+\pi^+\pi^-$ decays}

The high-statistics measurements of the $K\to 3\pi$ Dalitz Plot 
distributions performed by the NA48/2 collaboration~\cite{NA48a,NA48b}
have recently received a considerable 
attention because of the possibility to extract 
a precise information on  $\pi\pi$ scattering 
lengths~\cite{Cabibbo:2004gq,CI,Colangelo:2006va,Gamiz:2006km,Gevorkyan:2007ki}.

The leading mechanism which allow to measure 
$\pi\pi$ scattering lengths (and particularly the 
$a_{0}-a_{2}$ combination) in $K\to 3\pi$ decays is 
the $\pi^+\pi^- \to \pi^0\pi^0$  re-scattering at the $\pi^+\pi^-$
threshold, which produces a prominent cusp in the
$M_{\pi^{0}\pi^{0}}$ distribution of the 
$K^{+}\rightarrow\pi^{+}\pi^{0}\pi^{0}$ decay~\cite{Cabibbo:2004gq}.
The strength of this singularity is proportional to 
$a_{0}-a_{2}$, but also to phenomenological parameters 
introduced to describe the $K^+ \to \pi^+ \pi^+\pi^-$
amplitude (see Ref.~\cite{CI,Colangelo:2006va}). 
The latter must be determined by experiments from a fit 
to the $K^+ \to \pi^+ \pi^+ \pi^-$ decay distribution,
which is likely to receive sizable electromagnetic 
distortions because of the three 
charged particles in the final state.

\begin{figure}[t]
\begin{center}
\hspace{-0.3 cm}
\includegraphics[scale=0.46,angle=-90]{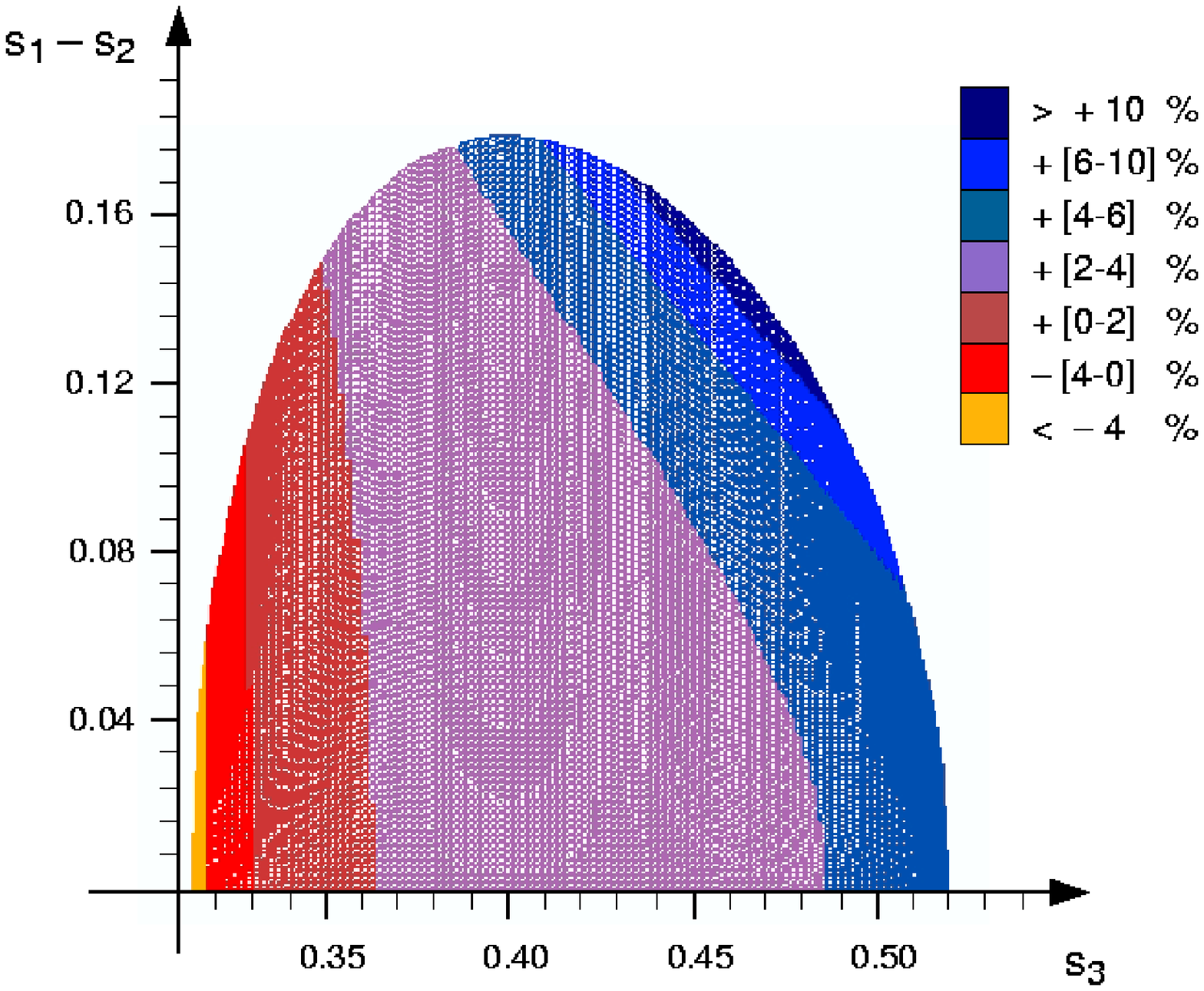}
\hspace{0.1 cm}
\includegraphics[scale=0.46,angle=-90]{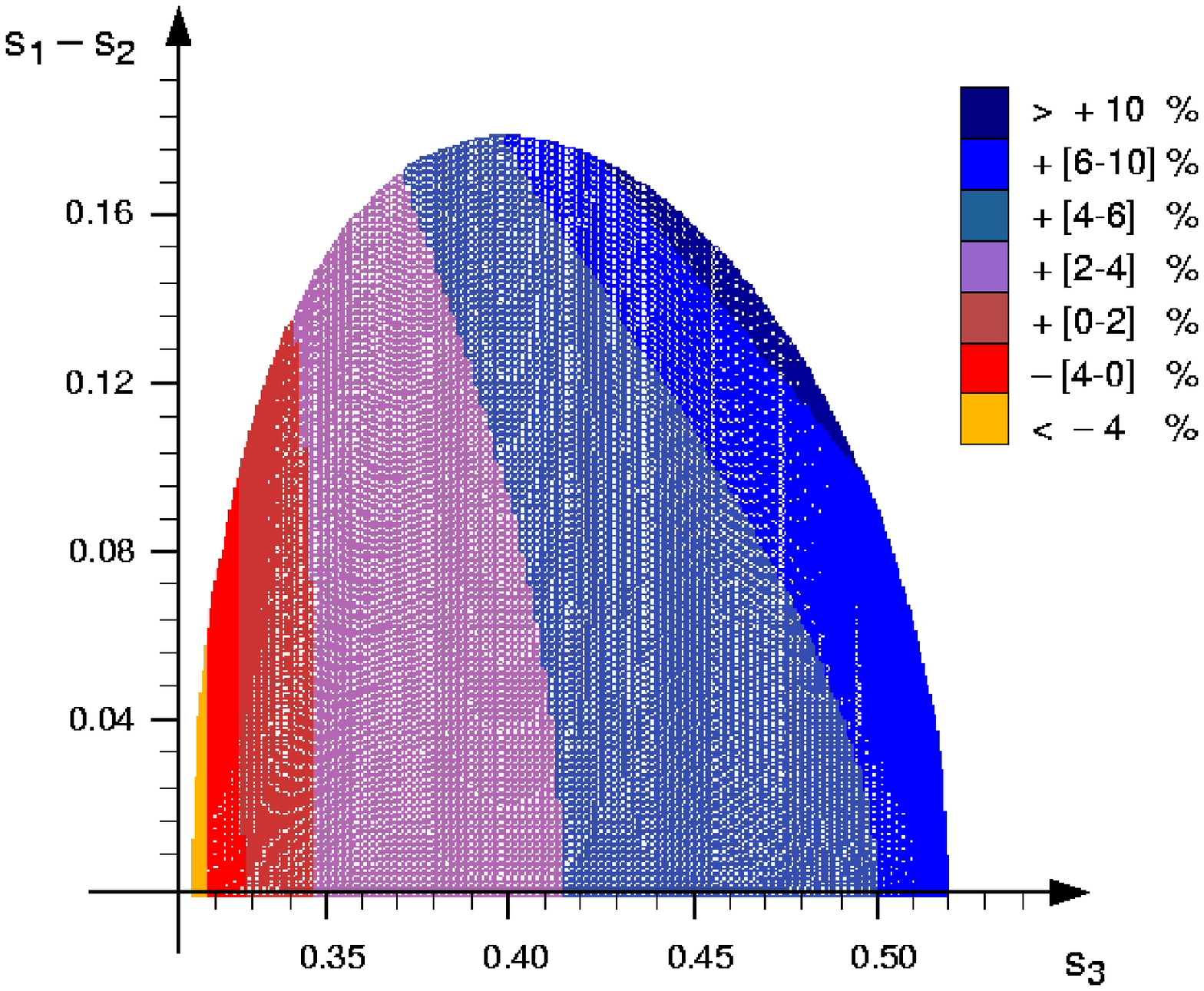}
\hspace{-0.3 cm}\\
\caption{\label{fig:K3p} Radiative corrections in the $K^-\to\pi^+\pi^+\pi^-$
decay. Left: density plot of $[\GG_{\rm eff}(s_{ij}; E)-1]$, 
evaluated with the full correction term in Eq.~(\ref{eq:main}) 
with $E= 5$~MeV.  Right: density plot of $[\GG_{\rm C}(s_{ij})-1]$
(Coulomb term only). The $s_i$ are in units of $m_K$. }
\end{center}
\end{figure}

In Fig.~\ref{fig:K3p} we show the impact of soft-photon corrections 
in the $K^-\to\pi^+\pi^+\pi^-$ decay distribution. In particular, 
we compare the result obtained with the full 
universal corrections factor in Eq.~(\ref{eq:main}) or
using only the Coulomb term in Eq.~(\ref{eq:Omega_C}).
As expected, radiative corrections induce sizable distortions,
especially at the border of the Dalitz plot distribution. 
However, these are well described by the Coulomb term up 
an overall normalization factor of $\cO(1\%)$.
The procedure adopted by the NA8/2 Collaboration 
to correct  $K^+ \to \pi^+ \pi^+ \pi^-$  data using only 
the Coulomb term is therefore well justified a posteriori. 

As discussed in the previous section, our 
general treatment do note take into account the 
formation of Coulomb bound states. Such process occur 
at the border of the $K^+ \to \pi^+ \pi^+ \pi^-$ Dalitz plot, 
when one of the two $\pi^+ \pi^-$ pairs is at rest.  
In order to determine the $K^+ \to \pi^+ \pi^+ \pi^-$
decay parameters relevant to the analysis of 
Ref.~\cite{CI,Colangelo:2006va}, 
the narrow regions at the border of the Dalitz plot 
with Coulomb corrections of  $\cO(100\%)$ 
should therefore be eliminated
by appropriate kinematical cuts. This procedure 
is perfectly consistent with the cut of the 
pionium region (around the peak 
of the $M_{\pi^{0}\pi^{0}}$ cusp) performed in 
Ref.~\cite{NA48a}.

\section*{Acknowledgments}
We thank Italo Mannelli for interesting discussions which initiated this 
analysis. This work is supported in part 
by the EU Contract No.~MRTN-CT-2006-035482 {\em FLAVIAnet}.

\end{document}